\documentclass{svjour2}                    
\smartqed  
\usepackage{graphicx}

\usepackage{latexsym}
\newcommand{\axj}{AXJ1838.0-0655}
\begin{document}

\title{The INTEGRAL - HESS/MAGIC connection:
a new class of cosmic high energy accelerators from keV to TeV
}


\author{   Pietro Ubertini on behalf of the IBIS Survey Team 
}


\institute{P. Ubertini         \at
             Istituto di Astrofisica Spaziale e Fisica Cosmica (IASF), Roma, Italy \\
              Tel.: +39 06 49934090\\
              Fax: +39 06 49934094\\
              \email{pietro.ubertini@iasf-roma.inaf.it}             \\
}

\date{Received: date / Accepted: date}

\maketitle

\begin{abstract}
The recent completion and operation of the High Energy
Stereoscopic System \cite{aharonian05}, an array of ground based
imaging Cherenkov telescopes, has provided a survey with
unprecedented sensitivity of the inner part of the Galaxy and
revealed a new population of very high energy gamma-rays sources
emitting at E$>$100 GeV. Most of them were reported to have no
known radio or X-ray counterpart and hypothesised to be
representative of a new class of dark nucleonic cosmic sources. In
fact, very high energy gamma-rays with energies E $>$ 10$^{11}$ eV
are the best proof of non-thermal processes in the universe and
provide a direct in-site view of matter-radiation interaction at
energies by far greater than producible in ground accelerators. At
lower energy INTEGRAL  has regularly observed the entire galactic
plane during the first 1000 day in orbit providing a survey in the
20-100 keV range resulted in a soft gamma-ray sky populated with
more than 200 sources, most of them being galactic binaries,
either BHC or NS \cite{bird}. Very recently, the INTEGRAL new
source IGR J18135-1751 has been identified as the soft gamma-ray
counterpart of HESS J1813-178 \cite{ube05} and AXJ1838.0-0655 as
the X/gamma-ray counterpart of HESS J1837-069 \cite{malizia}.

Detection of non thermal radio, X and gamma-ray emission from
these TeV sources is very important to discriminate between
various emitting scenarios and, in turn, to fully understand their
nature.

The implications of these new findings in the high energy Galactic
population will be addressed.

\end{abstract}

\section{Introduction}
HESS (High Energy Stereoscopic System), a ground-based Cerenkov
array telescopes has been operated since a few years and in 2004
has performed the first Galactic plane scan with a sensitivity of
a few percent of the Crab at energies above 100 GeV, resulting in
the discovery of eight sources, most of which without any
counterpart at different energies \cite{aharonian05,aharonian06}.
Particular attention was devoted to HESS J1813-178, not identified
with any known X/gamma ray emitter and hypothesised to be a dark
particle accelerator. Independently, and at the same time,
INTEGRAL discovered a new soft gamma-ray source, namely IGR
J18135-1751, identified as the counterpart of HESS J1813-178. This
high energy emitter, whose nature was still mysterious at the time
of the discovery was then associated with the supernova remnant
(SNR) G12.82$\_$0.02 \cite{ube05,brogan,helfand}. Even if a chance
coincidence cannot be completely ruled out in view of the 2 arcmin
INTEGRAL error box and the possible angular extension in the high
energy, the overall characteristics of this star forming region
\cite{churchwell} comprising SNR G12.82$\_$0.02 are consistent
with supernova/plerion origin. More recently, the MAGIC (Major
Atmospheric Gamma Imaging Cerenkov telescope) collaboration has
reported a positive observations of HESS J1813-178, resulting in a
gamma-ray flux consistent with the previous HESS detection and
showing a hard power law with $\alpha$ = 2.1 in the range from
0.4-10 TeV \cite{albert}.\\The detection of a substantial number
of very high energy Galactic sources emitting a large fraction of
energy in the GeV to TeV range has opened a new space window for
astrophysical studies related to cosmic particle acceleration.
Different types of Galactic sources are known to be cosmic
particle accelerators and potential sources of high energy gamma
rays: isolated pulsars/pulsar wind nebulae (PWN), Supernova
remnants, star forming regions, binary systems with a collapsed
object like a microquasar or a pulsar etc. The HESS detection of
several TeV emitters without any counterpart at different energy
has made the detection of X and gamma-ray emission from these
sources a key issue to disentangle the mechanisms active in the
different emitting regions and, in turn, to understand the source
nature. The IBIS gamma-ray imager on board INTEGRAL is a powerful
tool to search for their counterpart above 20 keV in view of the
arcmin Point Source Location Accuracy associated to $\sim$ mCrab
sensitivity for exposure $>$1Ms \cite{ubertini03}.

\begin{figure}
\centering
\includegraphics[height=5.5cm]{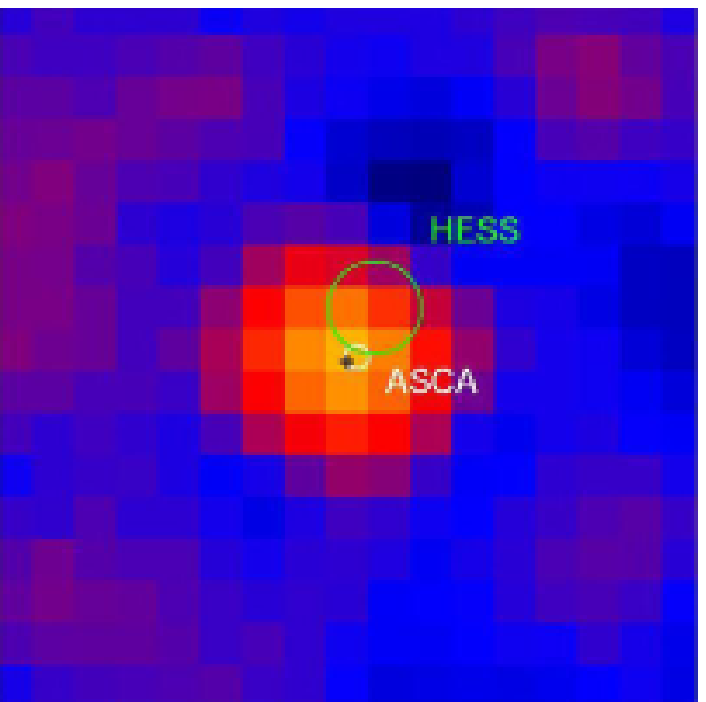}
\includegraphics[height=5.5cm]{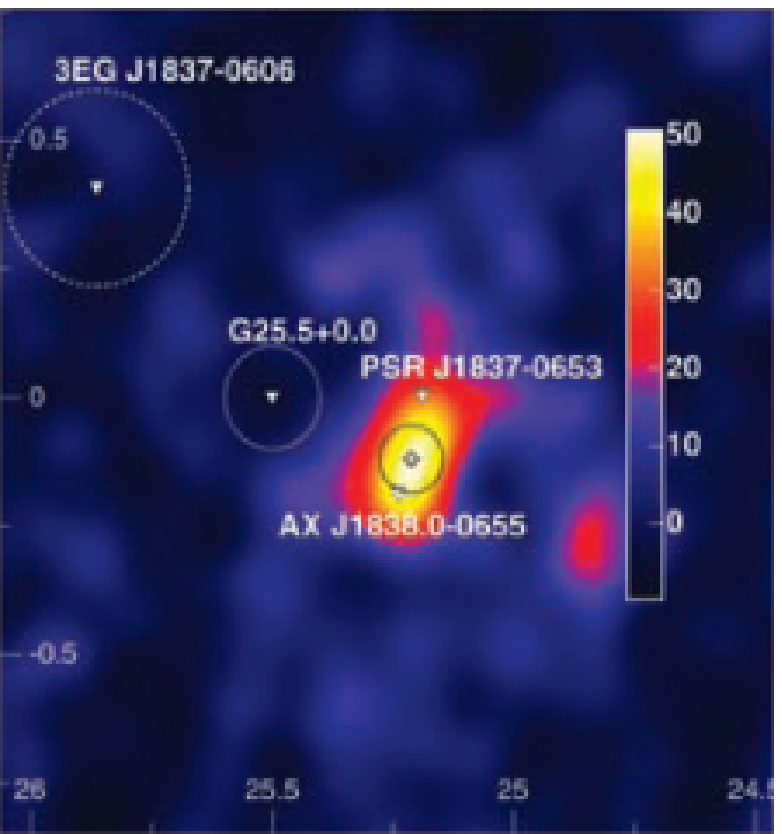}
\caption{Left panel: from Malizia et al. 2005. The IBIS/ISGRI
20-300 keV significance map showing the location of AX
J1838.0-0655 as well as the position and extension of HESS
J1837-069 (white circle) and the Einstein position (black cross).
The position and uncertainty of the ASCA source is basically
coincident with the central, brightest IBIS pixel. Right panel:
from Aharonian et al. 2005. The emission region of HESS J1837-069
overlapped to the ASCA error box of AXJ1838.0-0655. As can be seen
even if the two sources are positionally coincident the TeV
emission is suggestive of an extended object. } \label{fig:1}
\end{figure}

\begin{figure}
\centering
\includegraphics[height=7cm]{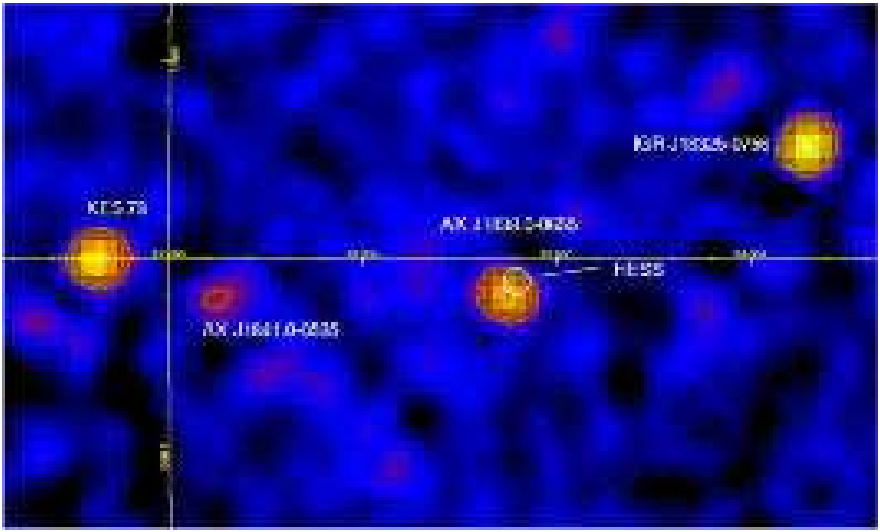}
\caption{The large IBIS sky region map containing AXJ1838.0-0655}
\label{fig:2}       
\end{figure}

\begin{figure}
\centering
\includegraphics[height=7.5cm]{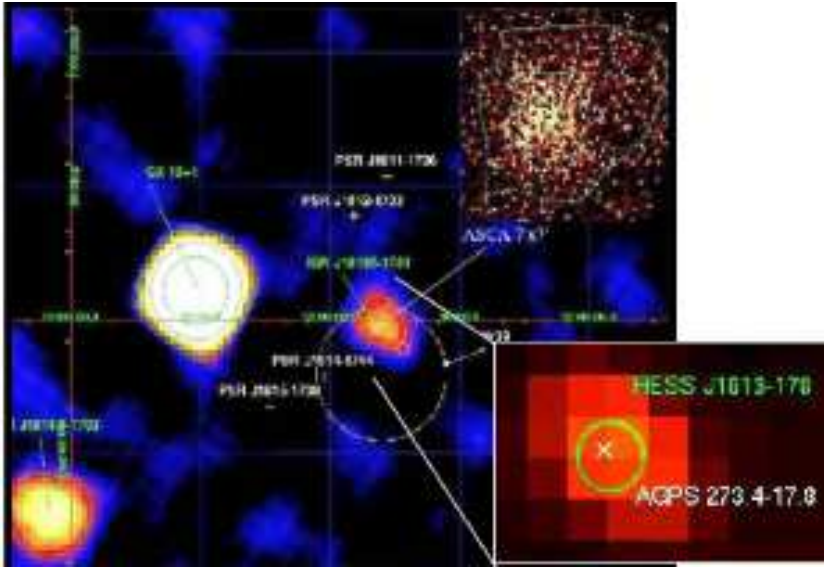}
\caption{From Ubertini et al. 2005. The IBIS/ISGRI 20-40 keV
significance map showing the location of IGR J18135- 1751 and
relative significance contours; the source spatial profile is
compatible with the detector response to a point source. The
extension of HESS J1813-178 as well as the position of
AGPS273.4-17.8 are both contained within the internal IBIS/ISGRI
contour. Also shown is the location (and extension) of W33, of the
4 nearest radio pulsars (PSR J1814- 1744,PSR J1812-1733, PSR
J1815-1738 and PSR J1811-1736) and of the ROSAT source 1WGA
J1813.7-1755 (white small diamond). The ASCA-SIS image is shown as
an insert on the top right side of the figure: contour levels
provide marginal evidence of extended emission. In the picture is
also present GX13+1 and the transient source SAX J1818.6-1703 that
are field sources not contaminating in any way IGR J18135-1751, in
view of the large angular distance between the objects (see text
for details). The coordinates are displayed in Galactic system
\cite{ube05} }
\label{fig:3}       
\end{figure}

\begin{figure}
\centering
\includegraphics[height=6.5cm]{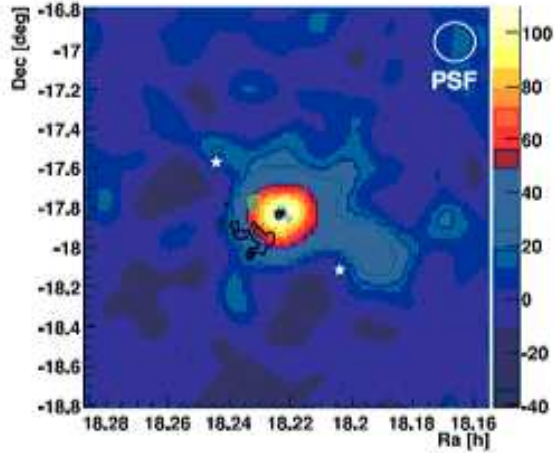}
\caption{From Albert et al. 2006. Sky map of gamma-ray candidate
events (background-subtracted) in the direction of HESS J1813-178
for an energy threshold of about 1 TeV. Overlaid are contours of
90 cm VLA radio (black) and ASCA X-ray data (green) from Brogan et
al. (2005). The two white stars denote the tracking positions W1,
W2 in the wobble mode (see \cite{albert} for details).}
\label{fig:4}       
\end{figure}

\begin{figure}
\centering
\includegraphics[height=7cm]{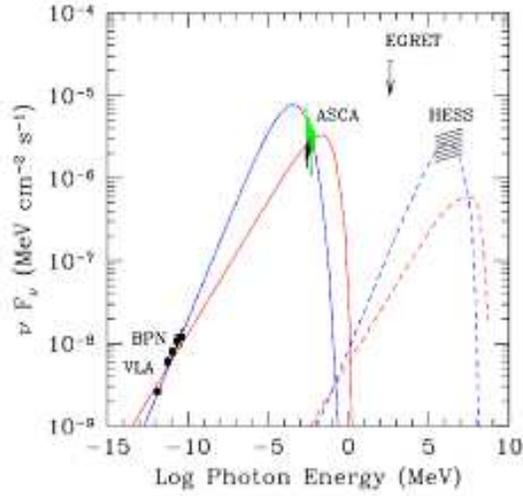}
\caption{From Brogan et al. 2005. Fits to the broadband emission
of HESS/ASCA source assuming that all the flux originates from the
shell of SNR G12.8 0.0. The diagonal black lines indicate both the
uncertainty in the HESS flux measurements, and the fact that no
spectral information has yet been published for the TeV emission.
The two models indicated by the red and blue lines show the range
of parameter space that best fit the data: the red model uses the
spectral index from the best fit ASCA Nh of 10.8$\times$10$^{22}$
cm$^{-2}$ (black X-ray spectrum), while the blue model uses the
spectral index implied by the 1$\sigma$ lower limit to Nh of 8.9
$\times$10$^{22}$ cm$^{-2}$ (green X-ray spectrum). Both models
include contributions from synchrotron (solid lines) and IC
(dashed lines) mechanisms. The authors have assumed that the
filling factor of the magnetic field in the IC emitting region is
15\%.}
\label{fig:5}       
\end{figure}

\begin{figure}
\centering
\includegraphics[height=6.5cm]{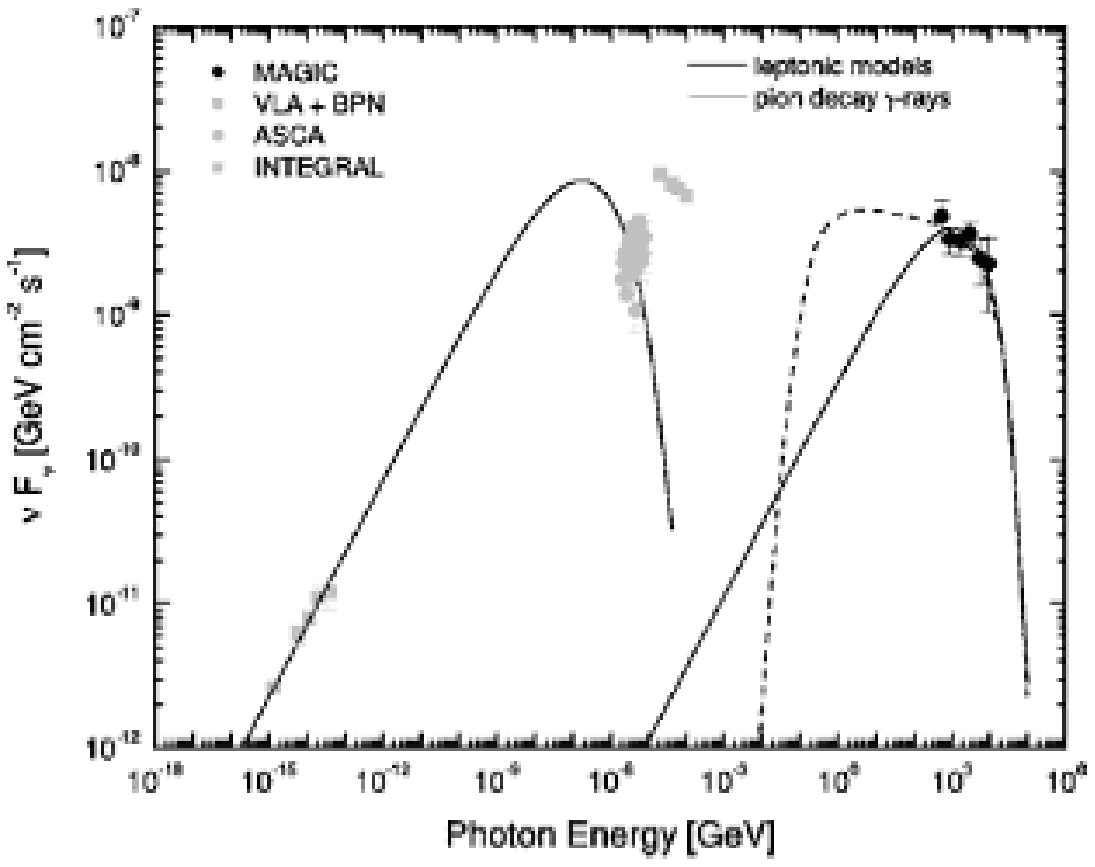}
\caption{From Albert et al. 2006. Leptonic and hadronic models for
the J1813-178 data. Details are given in \cite{albert}. Radio data
are from the VLA, Bonn, Parkes, and Nobeyama observatories
\cite{brogan}; X-ray and hard X-ray data are from ASCA
\cite{brogan} and INTEGRAL \cite{ube05}}
\label{fig:6}       
\end{figure}

\section{HESS J1837-069=AXJ1838.0-0655: the first IBIS/HESS new source}
\axj   is located in the Scutum arm region and has been detected
as a new INTEGRAL source by \cite{molkov} and \cite{bird} and it
is detected by IBIS up to 300 keV with a high statistical
significance, exceeding 15$\sigma$. The best positional location
is R.A.(2000) = 18h 38m 01.7s and Dec = -06$^{\circ}$ 54' 14.4''
with an error radius uncertainty of $\leq$3' \cite{malizia}. The
IBIS data provide a good fit with a simple power law with a
resulting $\chi^{2}_{\nu}/dof$=0.3/5 with  a photon index
$\Gamma$=1.66$\pm$0.23 (90\% c.l) and  a 20-300 keV flux of 9
$\times$ 10$^{-11}$ erg cm$^{-2}$ s$^{-1}$ \cite{malizia}. HESS
J1837-069 is one of the 10 very high energy sources located in the
central part of the Galaxy plane, detected in the TeV range with a
statistical significance of 7-8 $\sigma$. It is located at
R.A.(2000)= 18h 37m 42.7s and Dec(2000)=-06 55 39 (error box of
about 1-2') \cite{aharonian05} and the estimated flux, above 200
GeV, is 9$\times$ 10$^{-12}$ photons cm$^{-2}$ s$^{-1}$. The
authors suggest \axj  to be the candidate for HESS J1837-069, in
view of the spatial coincidence.

\axj    was discovered by the Einstein satellite in X-rays and
named 1E1835.3-0658 \cite{hertz} then observed by ASCA at higher
energy during the Galactic plane survey \cite{bamba} with a
positional uncertainty of 1' in radius, basically coincident with
the position of the original discovery. It was found to be bright
in the 0.7-10 keV band at a flux level of 1.1 $\times$ 10$^{-11}$
erg cm$^{-2}$ s$^{-1}$ and a hard ($\Gamma$=0.8) and absorbed
(N$_{H}$=4$\times$10$^{22}$ cm$^{-2}$) power law shape. The
summary of the observed positions for the different component,
from soft X-ray to TeV range, is shown in Fig. \ref{fig:1} (see
\cite{malizia} for details). From the angular distribution it
looks evident that the Einstein, ASCA and IBIS sources are the
same emitting source which is also likely to be active at TeV
energies. Finally, \cite{malizia} report the presence of a strong
radio source, TXS1835-069-CUL1835-06, associated in Simbad with a
candidate supernova remnant, SNR025.3-00.1 \cite{dulk}, positioned
at the side of the IBIS error box, though not compatible with
ASCA/Einstein X-ray position. The X/gamma-ray data indicate a
point source nature for this object, while the high energy results
are more in line with an extended emission, (see Fig. \ref{fig:1})
suggestive of a non-thermal radiation mechanism pointing to a SNR
and/or a PWN \cite{aharonian05}

\section{HESS J1813-178 and IGR J18135-1751: same emitting region/mechanism?}

IGR J18135-1751 was one of the newly discovered source during the
compilation of the second IBIS/ISGRI survey catalogue \cite{bird}.
It was immediately clear that the soft gamma-ray excess was
positionally coincident with the very high energy source named
HESS J1813-178, one of the 8 unknown sources found in the HESS
survey of the inner region of the Galactic plane. The source
position was found at R.A.(2000)=18h 13m 37.9s and
Dec(2000)=-17$^{\circ}$ 50' 34'' and has a positional uncertainty
of 1-2 arcmin. The source was reported to be slightly extended,
about 3 arcmin and had a statistical significance of about
9$\sigma$. Even if the source was reported to be quite bright 12
$\times$ 10$^{-12}$ photons cm$^{-2}$ s$^{-1}$ above 200 GeV,  it
was impossible to find an evident counterpart. An archival search
for soft X-ray data resulted in a possible counterpart in the ASCA
archive data: AGPS273.4-17.8 at R.A.(2000)=18h 13m 35.8s and
Dec(2000) = -17$^{\circ}$ 49' 43.35'' with an associated
uncertainty of 1'. In the X-ray band the source is fairly bright
showing a 2-10 keV flux (corrected for absorption) of 1.8 $\times$
10$^{-11}$ erg cm$^{-2}$ s$^{-1}$ \cite{ube05}.

In Fig. \ref{fig:3} is shown the IBIS/ISGRI 20-40 keV map with the
location of IGR J18135-1751. The authors report the relative
significance contours to be 6 (for the external one), 8, 10, 20
and 40 $\sigma$, that are compatible with a point source. The
extension of HESS J1813-178 as well as the position of
AGPS273.4-17.8 are both contained within the internal IBIS/ISGRI
contour. Also it is shown the location and the extension of W33
and the 4 nearest radio pulsars, namely PSR J1814-1744, PSR
J1812-1733, PSR J1815-1738 and PSR J1811-1736. The combined
IBIS-ASCA spectrum, obtained adding the two data set without the
need of any normalisation, confirm that HESS J1813-178 has a point
like X-ray counterpart with a power law emission from 2 to 100 keV
and an associated radio counterpart. The data set is strongly
suggesting that it is a non-thermal source, possibly accelerating
electrons and positrons which radiate through synchrotron and
inverse Compton mechanism. Brogan et. al. (2005) have provided a
possible interpretation of the HESS/ASCA spectrum by fitting the
broadband emission assuming that all the flux originates from the
shell of SNR G12.8 (see Fig. \ref{fig:5} and \cite{brogan}, for a
detailed description). The two proposed models (corresponding to
two different Nh absorption values) are shown by the red and blue
lines that provide the best fit to the data. The models include
X-ray emission from synchrotron radiation (solid lines) and
Inverse Compton processes (dashed lines), assuming a filling
factor of 15\% for the magnetic field in the IC emitting region.
More recently the MAGIC experiment has observed HESS J1813-178,
resulting in the detection of a differential gamma-ray flux
consistent with a hard-slope power law, described as \linebreak
$\mathrm{d}N_{\gamma}/(\mathrm{d}A \mathrm{d}t \mathrm{d}E) = (3.3
\pm 0.5) \times 10^{-12} (E/\mathrm{TeV})^{-2.1 \pm 0.2} \
\mathrm{cm}^{-2}\mathrm{s}^{-1} \mathrm{TeV}^{-1}$ \cite{albert}.
The image of the MAGIC field containing the high energy excess is
shown in Fig. \ref{fig:4}. The authors quote the systematic error
to be 35\% in the flux level determination and 0.2 for the
spectral index. Within errors, the flux was found steady in the
timescales of weeks as well as in the year-long time span between
the MAGIC and HESS pointings. They report a multiwavelength
emission associated to HESS J1813-178, is shown in Fig.
\ref{fig:6}, including the MAGIC data at high energies. The
authors compare the hadronic and leptonic emission models with the
high-energy gamma-ray data (see \cite{albert} and \cite{torres}
for details). The main conclusions are that for hadronic models
the observed high energy luminosity (2.5$\times 10 ^{34} erg
s^{-1}$ at a distance of 4kpc) implies a matter density of $\sim$6
cm$^{-3}$ assuming gamma-ray are generated in the whole Supernova
remnant, an acceleration efficiency for the hadrons of $\sim$3\%
and a Supernova power of 10$^{51}$  ergs with a target mass for
relativistic particles of about 2 Solar masses (within or close to
the SNR) to justify the observed luminosity. \\For leptonic
models, they assume relativistic electrons distribution of
\linebreak $\mathrm{d}N_{e}/(\mathrm{d}V \mathrm{d}E)=A_e (E/{\rm
GeV})^{-\alpha_e} \exp{(-E/E_{\rm max, e})}\; {\rm GeV^{-1}
cm^{-3}} $ and obtain good fits with value of $\alpha_e$
$\sim$2.0-2.1 $E_{\rm max, e}$ $\sim$20-30 TeV, assuming the
cosmic microwave background as target photons. The best fit to the
radio synchrotron emission is $\alpha$ $\sim$ 2.0 and require a
magnetic field of 10 $\mu$G with a filling fraction of about 20\%
(\cite{albert}). This model is not very different from the blu one
in Brogan et al. (see Fig. \ref{fig:5}), even if a lower filling
factor $E_{\rm max, e}$ is assumed.

\section{Implication of the Synchrotron - Inverse Compton Scenario}

The detection of soft gamma ray photons from TeV sources it is
important to better understand the emission mechanisms and
particle acceleration processes in SNR. In addition, the
consistency with a power law of X-ray spectra, spanning from few
to few hundred of keV, and the lack of X/gamma variability is
compatible with a SNR or a Pulsar Wind Nebula, the latter not yet
evident fom observational view point. As an example, HESS
J1813-1751 has a very young remnant, of the order of 300 to 3000
year, assuming a density medium of $\sim$1 cm$^{-3}$. The IBIS
detection of soft gamma ray photons up to $\sim$100 keV
\cite{ubertini06} in the Synchrotron - Inverse Compton Scenario
force the lifetime of the emitting electrons to be quite short if
compared to the one of the radiating electrons in the GHz
frequency \cite{lazendic,con98,ubertini06}. In fact, the lifetime
for the high energy electrons is $t_{1/2} \sim 224 \times
(B/10{\mu}G)^{-3/2}$ y. This model also imply the radio and X/soft
gamma-ray synchrotron emission to be spatially coincident and
X-ray emission that sharply drops behind the shock. Because of
this effect, the X-ray photons will be confined close to the
electrons acceleration region while the same electron population
would eventually drift having a diffusion speed of a fraction of
the light one. The detection of a substantial flux of soft
gamma-rays from the HESS J1813-1751 region  is supportive of the
red fit model proposed by \cite{brogan}, that implies a lower
ratio of peak emission frequencies $R_\nu
=\nu_{p}^{IC}/\nu_{p}^{Sy}$ if compared with the blue one, not
capable to fit the IBIS data and, in turn, not compatible with the
single source scenario. Unfortunately, the red model is unable to
fit the HESS data, the physical reason being the synchrotron
losses, in the presence of a quite strong magnetic field predicted
by the model, constraining the density of the electrons necessary
to radiate via Inverse Compton interaction with the CMB radiation.
The picture ameliorate considering that a substantially higher UV
light flux could be supplied by W33, a close HII region. An energy
density of $\sim$3 eV cm$^{-3}$ could
 be easily provided, a factor $\sim$10 higher than the density of
 $\sim$0.26 eV cm$^{-3}$ of the CMB photons \cite{aharonian05}
(for a detailed analysis of the IC model see \cite{ubertini06}).

\section{Conclusions}
\begin{figure}
\centering
\includegraphics[height=6.5cm]{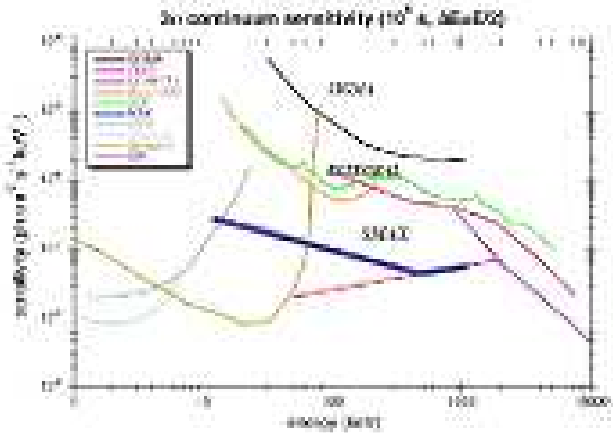}
\caption{The lap in sensitivity expected from new generation gamma
ray observatorysuch as MAX+ or the newly prosed Gamma Ray Imager
(GRI) based on Laue Lens complemented at lower energy by a large
CZD detrector (in the figure is shown the SMAX  sensitivity, this
conference)}
\label{fig:7}       
\end{figure}

It is clear that both the hadronic and leptonic models so far
proposed fails to easily explain the whole observational picture
if the radio, X/soft gamma-rays and TeV high energy photons are
produced in the same SNR region by a single physical process. In
fact, to finally confirm the above hypothesis it is necessary to
have instruments capable to provide spatially resolved
spectroscopy with a fraction of arcsec. While in the X-ray range
this seems to be possible with long CHANDRA exposures it is not
with the present generation of gamma ray instruments in the best
case providing arcmin angular resolution \cite{lebrun,ubertini03}.
This will be achievable with a new generation of Gamma Ray Imagers
providing at the same time a factor of 10 better sensitivity, as
discussed in the present conference, as shown in Fig. \ref{fig:7}.
On a shorter time frame  it could be possible to improve the lack
of our knowledge in the 'blank region' in between the soft gamma
ray range covered by INTEGRAL and the very high energy one by HESS
and MAGIC. In fact, the imminent launch of the Italian AGILE
gamma-ray satellite, covering the 30 MeV-50 GeV range with a good
sensitivity over a wide field of view, and GLAST in perspective,
will be a powerful tool to finally disentangle the nature of the
high energy sky.

\begin{acknowledgements}
The author acknowledge this research has been granted by the
Italian Space Agency via contract n. I/R/046/04 ASI/IASF.
\end{acknowledgements}

\begin{thebibliography}{}

\bibitem{aharonian05} Aharonian, F. et al., {\em Science} (2005), {\bf 307}, 1938
\bibitem{aharonian06} Aharonian, F.,  et al., {\em ApJ}, (2006), {\bf 636}, 777
\bibitem{albert} Albert, J.  et al., {\em ApJ}, (2006), {\bf 637}, L41
\bibitem{bamba} Bamba, A. et al., {\em ApJ}, (2003), {\bf 589}, 253
\bibitem{bird}Bird A.J., {\em ApJ}, (2006), {\bf 636}, 765
\bibitem{brogan} Brogan, C. et al., {\em ApJ}, (2005), {\bf
629}, L105
\bibitem{churchwell} Churchwell, E. , {\em A\&A Rev.}, (1990) {\bf 2}, 79
\bibitem{con98} Condon, J.J. et al.,  {\em AJ}, (1998), {\bf 115}, 1693
\bibitem{dulk} Dulk, P. $\&$ O. Slee, {\em AJPh}, (1972), {\bf 25}, 429
\bibitem{helfand} Helfand, D. J.Becker, R. H., \& White, R. L. (2005), {\bf astroph/
0505392}
\bibitem{hertz} Hertz P. \& Grindlay J. {\em AJ }, (1998), {\bf 96}
\bibitem{lazendic} Lazendic, J.S et al., {\em ApJ}, (2004), {\bf 602}, L271
\bibitem{lebrun} Lebrun et al., {\em A\&A}, (2003), {\bf 411}, L141
\bibitem{malizia} Malizia, A. {\em ApJ}, (2006), {\bf 630}, L157
\bibitem{molkov} Molkov S. et al.,  {\em Astr. Lett.}, (2004), {\bf 30}, 534
\bibitem{torres} Torres D.F. et al.,  {\em Phys. Rept}, (2003), {\bf 382}
303
\bibitem{ubertini03} Ubertini, P. et al., {\em A\&A}, (2003), {\bf 411}, L131
\bibitem{ube05} Ubertini, P. et al., {\em ApJ}, (2005), {\bf 629}, L109
\bibitem{ubertini06} Ubertini, P. et al., {\em Proceedings - Conference "A Life with Stars"
         Amsterdam 22-26/August 2005} - {\bf Elsevier}

\end{thebibliography}
\end{document}